\documentclass[pre,showpacs,amsfonts,amssymb,eqsecnum,%
twocolumn,floatfix,superscriptaddress]{revtex4}
\usepackage{amsmath}
\usepackage{graphicx}
\usepackage{bm}
\allowdisplaybreaks
\usepackage{epstopdf}
\usepackage{epsfig}

\begin{document}

\title{Stochastic dynamics and logistic population growth}

\author{Vicen\c{c} M\'{e}ndez}
\affiliation{Grup de F{\'i}sica Estad\'{i}stica.  Departament de F{\'i}sica.
Facultat de Ci{\`e}ncies. Edifici Cc. Universitat Aut\`{o}noma de Barcelona,
08193 Bellaterra (Barcelona) Spain}
\author{Michael Assaf}
\affiliation{Racah Institute of Physics, Hebrew University of Jerusalem,
Jerusalem 91904, Israel}
\author{Daniel Campos}
\affiliation{Grup de F{\'i}sica Estad\'{i}stica.  Departament de F{\'i}sica.
Facultat de Ci\`encies. Edifici Cc. Universitat Aut\`{o}noma de Barcelona,
08193 Bellaterra (Barcelona) Spain}
\author{Werner Horsthemke}
\affiliation{Department of Chemistry, Southern Methodist University,
Dallas, Texas 75275-0314, USA}
\date{\today}

\begin{abstract}
The Verhulst model is probably the best known macroscopic rate equation in
population ecology. It depends on two parameters, the intrinsic growth rate
and the carrying capacity. These parameters can be estimated for different
populations and
are related to the reproductive fitness and the competition for limited
resources, respectively. We investigate analytically and numerically the
simplest possible microscopic scenarios that give rise to the logistic
equation in the deterministic mean-field limit. We provide a definition of
the two parameters of the Verhulst equation in terms of microscopic
parameters. In addition, we derive
the conditions for extinction or persistence of the population by employing
either the ``momentum-space'' spectral theory or the ``real-space'' Wentzel-
Kramers-Brillouin (WKB) approximation to determine the probability
distribution function and the mean time to extinction of the population. Our
analytical results agree well with numerical simulations.
\end{abstract}

\maketitle

\section{Introduction}
Quantitative models of population dynamics have attracted an enormous
interest from biology to mathematics and physics
\cite{MeCaBa14,LJSA10,Re93}. In the deterministic limit, these models
coincide with macroscopic rate equations based on phenomenological laws.
The simplest one corresponds to Malthus law, where the per capita rate of
change in the number of individuals is constant, resulting in a linear
growth rate for the population, $dn/dt=rn$. The population grows
exponentially, $n(t)=n_{0}\exp(rt)$, where $r$ is
the intrinsic growth rate and
$n_{0}$ is the initial population. Unlimited exponential growth
is patently unrealistic, and factors that regulate growth
must be taken into account.
The most famous extension of the exponential growth model is the
Verhulst model, also known as the logistic model, where the per capita
rate of change decreases linearly with the population size. The population's
growth rate, $dn/dt=rn(1-n/K)$, is now a quadratic function of the population
size, where $K$ is known as the carrying capacity. This equation was derived
initially by P. Verhulst in 1845 \cite{Ve45,Ve47} and was rediscovered later
by R. Pearl in 1920 \cite{Pe20}. Other models, like the Gompertz growth,
$dn/dt=\alpha n\ln(K/n)$, exhibit many of the same properties,
but the logistic
equation is arguably the best-known
and most widely applied rate equation for
population growth and population
invasion \cite{MeFeHo10,MeCaBa14}.

These models are deterministic and ignore fluctuations. Real
populations evolve in a stochastic manner, experiencing intrinsic
noise (or
internal fluctuations) caused by the discreteness of individuals
and the
stochastic nature of their interactions, see, e.g.,
\cite{Fe39,Ke49,GoRD74,Naa01,BjGr01,OvMe10}.
When the typical size of the
population is large, fluctuations in the observed number of
individuals are
typically small in the absence of external or environmental noise.
The dynamics of the population then can be
described by a deterministic mean-field rate equation. In the case of
the logistic equation, the population evolves from an initial condition
to a stable stationary state, where the population size equals the
carrying capacity and persists forever. However, if the
typical population
size is not large, internal fluctuations can lead to the extinction of
the population \cite{Ni77}. The effects
of internal fluctuations have been studied in predator-prey models
\cite{NiPr77,MKNe05},
epidemic models \cite{KrLe89,Naa99a,KeRo08,Scetal09,BrNe10,Br10,Sc11},
cell
biology \cite{Br14}, and ecological systems \cite{OvMe10}. In particular,
extinction of a stochastic population \cite{De91,Naa01,AsMe10}, which is a
crucial concern for population biology  \cite{MeHa08} and
epidemiology \cite{Es04,BlMc10}, has also attracted
scrutiny in cell biochemistry \cite{AsMe08} and in physics
\cite{VK92,Ga90}.

To describe the intrinsic noise of populations, we adopt
individual-based models, also called stochastic
single patch models \cite{BlMc12,ZeSoYa15}.
An individual-based formulation provides several advantages.
It is often easier to define an ecological system
in terms of the events that govern the dynamics of
the system at the level of individuals.
Population-level models, such as the Verhulst equation,
can then be derived analytically as
the mean-field approximation,
instead of simply be
postulated phenomenologically. In this way,
individual-based models provide a
microscopic basis for the usual ecological
rate equations,
and the range validity of the latter can be established
by comparing its predictions with those of
the former.
Individual-based models capture the fact that
populations consist of
discrete individuals undergoing
random events corresponding to
birth (reproduction of the population),
competition (between individuals for limited resources)
and death (natural decay
of individuals).
It is well known that different types of
of individual-based schemes
are described
by the same Verhulst equation in the
deterministic limit. Since extinction is ultimately
caused by the stochastic nature of the interactions
between individuals, it
is critical to analyze how the details of the individual
processes
affect the ultimate fate of the population
or the time to extinction.
We
explore a variety of stochastic interactions
between individuals,
all of which give rise to the logistic
equation in the mean-field limit. We
find different dynamical behaviors, such as persistence
or extinction, of a
population that experiences birth, death, and competition processes.
Extinction is due to rare fluctuations, and the
mean
extinction time (MET) of the population
strongly
depends on the microscopic details of the processes,
such as the number of ``newborn'' individuals
or the number of individuals removed due to exclusive competition.
We obtain
analytical solutions for the probability distribution function (PDF) of
individuals, if the population persists, and for the MET, if the
population becomes extinct.
Our analytical results are compared with numerical
simulations, performed using the first reaction method \cite{ToCo14}.
We consider the birth-and-death and birth-competition-death
cases separately,
making use of the ``momentum space" spectral
theory \cite{AsMe06, AsMe07}
and the ``real-space" WKB theory \cite{kessler2007extinction,meerson2008noise,escudero2009switching,AsMe10}, respectively.

\section{Master and mean field equations for
general birth-competition-death processes}

We investigate individual-based models
of populations in which the following birth,
competition, and death processes occur,
\begin{subequations}
\label{gsys}
\begin{align}
b\,\text{X} & \xrightarrow{\lambda}  (a+b)\,\text{X},\\
c\,\text{X} & \xrightarrow{\mu} (c-d)\,\text{X},\\
\text{X} & \xrightarrow{\gamma} \emptyset,
\end{align}
\end{subequations}
where $a$, $b$, $c$, and $d$  are positive integers, and $d\leq c$.
Such processes occur also in chemically
reacting systems, and
it is convenient to adopt the language of chemical
kinetics to make a connection with the literature of stochastic
chemical models.
Therefore, we will often refer to the processes of \eqref{gsys}
as ``reactions.''
If $d=c$, the last two reactions are death reactions, due to competition
between $c$ individuals ($cX \overset{\mu}{\rightarrow} \emptyset$)
or due to natural decay ($X  \overset{\gamma}{\rightarrow}  \emptyset$).
We make the standard assumption that the reaction scheme \eqref{gsys}
defines a Markovian
birth-and-death
process, see, e.g.,  \cite{NiOrDeRo74,NiPr77,Gi77a,Ga90},
and employ the Master equation,
also known as the forward Kolmogorov equation,
to describe the temporal evolution
of $P(n,t)$, the probability of having $n$
individuals at time $t$,
\begin{equation}
\frac{\partial P(n,t)}{\partial t}=\sum_{r}
\left[W(n-r,r)P(n-r,t)-W(n,r)P(n,t)\right].
\label{eq:me}
\end{equation}
Here, $W(n,r)$ are the transition rates
between the states with $n$ and $n+r$ individuals, and
$r=\{r_{1},r_{2},r_{3}\}=\{a,-d,-1\}$
are the transition increments.
Equation \eqref{eq:me} can generally only
be solved in the stationary limit,
$\partial P(n,t)/\partial t=0$
and only for
the special case that $a=d=1$, i.e., only single-step
processes occur in the population. Then the
condition of
detailed
balance holds, which significantly simplifies
the theoretical analysis, and exact analytical expressions
can in principle be obtained for the stationary PDF or
the MET. For recent reviews, see for
example \cite{Naa01, DoSaSa05,Na11a}. We emphasize that we
study the general generic case of arbitrary $a$ and $d$
to elucidate how the microscopic details affect the PDF
or the MET.
Calculating the stationary PDF or the MET is highly
nontrivial for multi-step reactions, and this case has only
recently began to be addressed.

The transition rates
corresponding to
each reaction, $W(n,r)$, are obtained
from the reaction kinetics \citep{Ga90} and for \eqref{gsys}
read:
\begin{subequations}
\label{trans}
\begin{align}
W(n,a)&=\frac{\lambda}{b!}\frac{n!}{(n-b)!},\\
W(n,-d)&=\frac{\mu}{c!}\frac{n!}{(n-c)!},\\
W(n,-1)&=\gamma n.
\end{align}
\end{subequations}
Substituting (\ref{trans}) into (\ref{eq:me}), we find
\begin{multline}
\frac{\partial P(n,t)}{\partial t}=\frac{\lambda}{b!}
\frac{(n-a)!}{(n-a-b)!}P(n-a,t)\\
+\frac{\mu}{c!}\frac{(n+d)!}{(n+d-c)!}P(n+d,t)+\gamma (n+1)P(n+1,t)\\
-\left[\frac{\lambda}{b!}\frac{n!}{(n-b)!}+\frac{\mu}{c!}
\frac{n!}{(n-c)!}+\gamma n\right]P(n,t),\label{eq:me2}
\end{multline}
where it is understood that $P(n<0,t)=0$.
The probability generating function \cite{Ga90}
is defined as
\begin{equation}
G(p,t)=\sum_{n=0}^{\infty}p^{n}P(n,t),
\label{eq:dG}
\end{equation}
where $p$ is an auxiliary variable, which is conjugate to
the number of particles \cite{KaEl04}. Once $G(p,t)$ is known,
the PDF is given by the Taylor coefficients
\begin{equation}
P(n,t)=\frac{1}{n!}\left[\frac{\partial^{n}G(p,t)}{
\partial p^{n}}\right]_{p=0}.
\label{Pnt}
\end{equation}

Normalization of $P(n,t)$
implies that $G(p=1,t)=1$. Multiplying (\ref{eq:me2}) by $p^{n}$,
summing over $n$, and renaming the index of summation, we find
\begin{multline}
\frac{\partial G(p,t)}{\partial t}=\frac{\lambda}{b!}
\sum_{n=0}^{\infty}(p^{n+a}-p^{n})\frac{n!}{(n-b)!}P(n,t)\\
+\frac{\mu}{c!}\sum_{n=0}^{\infty}(p^{n-d}-p^{n})\frac{n!}
{(n-c)!}P(n,t)\\
+\gamma \sum_{n=0}^{\infty}(p^{n-1}-p^{n})nP(n,t).
\label{eq:me3}
\end{multline}
Taking into account the property
\begin{align}
p^{k}\frac{\partial^{k}G(p,t)}{\partial p^{k}}=&
\sum_{n=0}^{\infty}n(n-1)\cdots(n-k+1)p^{n}P(n,t)\nonumber\\
=&\sum_{n=0}^{\infty}\frac{n!}{(n-k)!}p^{n}P(n,t)
\end{align}
in (\ref{eq:me3}), we finally obtain the evolution equation
for $G(p,t)$,
\begin{multline}
\frac{\partial G(p,t)}{\partial t}=\frac{\lambda}{b!}p^{b}
(p^{a}-1)\frac{\partial^{b}G}{\partial p^{b}}\\
+\frac{\mu}{c!}p^{c-d}(1-p^{d})\frac{\partial^{c}G}
{\partial p^{c}}
+\gamma (1-p)\frac{\partial G}{\partial p}.
\label{eq:G}
\end{multline}
Equation (\ref{eq:G}) is exact and equivalent to the Master equation
(\ref{eq:me2}). If only one individual reactant
is present in all the reactions,
i.e., $b=c=1$, then (\ref{eq:G}) is first order in $p$
and can be solved exactly using the method of characteristics.

Macroscopic equations, i.e., equations for the expected
or average values, can
be obtained easily from (\ref{eq:me2}). Multiplying (\ref{eq:me2})
by $n^{k}$, summing up over $n$, and renaming
the index of summation, we find
\begin{multline}
\frac{\partial}{\partial t}\sum_{n=0}^{\infty}n^{k}P(n,t)
=\\
\frac{\lambda}{b!}\sum_{n=0}^{\infty}\left[(n+a)^{k}-n^{k}\right]
\frac{n!}{(n-b)!}P(n,t)\\
+\frac{\mu}{c!}\sum_{n=0}^{\infty}\left[(n-d)^k-n^{k}\right]
\frac{n!}{(n-c)!}P(n,t)\\
+\gamma \sum_{n=0}^{\infty}\left[(n-1)^k-n^{k}\right]n P(n,t).
\end{multline}
The $k$-th moment is defined
as $\left\langle n^{k}\right\rangle =
\sum_{n=0}^{\infty}n^{k}P(n,t)$
and evolves according to the ordinary differential equation
\begin{multline}
\frac{d\left\langle n^{k}\right\rangle}{dt}=
\frac{\lambda}{b!}
\left\langle \left[(n+a)^{k}-n^{k}\right]
\prod_{m=0}^{b-1}(n-m)\right
\rangle\\
+\frac{\mu}{c!}\left\langle \left[(n-d)^{k}-n^{k}\right]
\prod_{m=0}^{c-1}
(n-m)\right\rangle\\
 +\gamma\left\langle \left[(n-1)^{k}-n^{k}\right]n\right\rangle.
\label{eq:moments}
\end{multline}
Equation (\ref{eq:moments}) is not closed, and one
must deal with a hierarchy of coupled differential equations
for $k=1,2,3,\dotsc$. In order
to truncate this set and to obtain closed equations,
we make use of the mean-field approximation
$\left\langle n^{k}\right\rangle \simeq\left\langle n
\right\rangle^{k}$,
which holds if the typical population size is
large \cite{VK92,Ga90}.
For $k=1$, the mean-field equation reads
\begin{equation}
\frac{d\rho}{dt}=\frac{\lambda a}{b!}\rho^{b}
-\frac{\mu d}{c!}\rho^{c}
-\gamma\rho,
\label{eq:mfe}
\end{equation}
where $\rho=\left\langle n\right\rangle$ is a macroscopic quantity,
the average or expected number of individuals in the population.

\section{Birth and death/competition processes}

We consider the case of two reactions, i.e., $\gamma=0$:
\begin{subequations}
\label{eq:bd}
\begin{align}
b\,\text{X}&\xrightarrow{\lambda}(b+a)\,\text{X},\\
c\,\text{X}&\xrightarrow{\mu}(c-d)\,\text{X}.
\end{align}
\end{subequations}
In the first reaction (birth), $b$ individuals have to
interact with each other to produce $a$ new individuals
at a constant
rate $\lambda$. In the second reaction (death by competition),
$c$ individuals
interact with each other to remove $d$ individuals at a constant
rate $\mu$. The fact that $b$, in general, can be larger
than $1$ includes scenarios where a single individual
cannot generate by itself
new individuals, which represents a type of Allee
effect \cite{St99}.

Equation (\ref{eq:mfe}) reduces to the logistic equation
if $b=1$ and
$c=2$. From a
kinetic point of view this means that an individual does not need to
interact to give rise to new individuals; the birth reaction takes
the form $\text{X}\xrightarrow{\lambda}(a+1)\,\text{X}$.
The fact $c=2$ implies that a linear death rate,
corresponding
to $\text{X}\rightarrow \emptyset$, cannot occur
for the scheme
\eqref{eq:bd} in this case.
The possible death reactions, compatible with a mean-field logistic
equation,
are $2\,\text{X}\xrightarrow{\mu}\text{X}$ (competition)
or $2\,\text{X}\xrightarrow{\mu}\emptyset$ (annihilation).
Consequently, the
birth-and-death processes that lead to logistic macroscopic behavior are
\begin{subequations}
\label{scheme1}
\begin{align}
\text{X}&\xrightarrow{\lambda}(a+1)\,\text{X},\\
2\,\text{X}&\xrightarrow{\mu}\text{X},
\end{align}
\end{subequations}
and
\begin{subequations}
\label{scheme2}
\begin{align}
\text{X}&\xrightarrow{\lambda}(a+1)\,\text{X},\\
2\,\text{X}&\xrightarrow{\mu}\emptyset.
\end{align}
\end{subequations}
The logistic equation for these two reaction schemes reads
\begin{equation}
\label{RElogistic}
\frac{d\rho}{dt}=r\rho\left(1-\frac{\rho}{N}\right),
\end{equation}
where
\begin{equation}
r\equiv a\lambda\,\text{and}\, N\equiv 2a\lambda/\mu d
\end{equation}
are the intrinsic growth rate and the carrying capacity,
respectively.
These definitions are valuable because they allow us to
relate the
macroscopic parameters $r$ and $N$, which can be
measured for different
kinds of populations, to the microscopic parameters
that characterize the stochastic
processes involved in the interaction between the
individuals of the
population.
From a macroscopic point of view, the logistic equation
for population growth is specified by two parameters.
On the other hand,
the schemes \eqref{scheme1} and \eqref{scheme2} contain
up to
four microscopic parameters, namely $a$, $d$, $\mu$ and $\lambda$.
As a result,
we have two additional free microscopic parameters that
can take arbitrary
positive values compatible with the same mean-field
logistic equation.
Rate equation (\ref{RElogistic}) has an unstable steady
state at $\rho_{s}=0$ and a
stable steady state at $\rho_{s}= N$ for
$d=1$ or $d=2$.
Below we
deal separately
with schemes (\ref{scheme1}) and (\ref{scheme2})
and apply the
momentum-space ($p$-space) spectral theory to find
the stationary PDF in the
case of population survival
or the MET in the case of population
extinction.
An important advantage of the $p$-space representation
stems from the fact
that
the evolution equation for the generating function
$G(p, t)$ is exactly
equivalent to the original master equation.
Therefore the $p$-space approach is especially valuable
for an exact analysis.

\subsection{Case I: $\text{\rm X}\xrightarrow{\lambda}(a+1)\,
\text{\rm X}$, $2\,\text{\rm X}\xrightarrow{\mu}
\text{\rm X}$}\label{subs:case1}

\begin{figure}[htbp]
\includegraphics[width=\hsize]{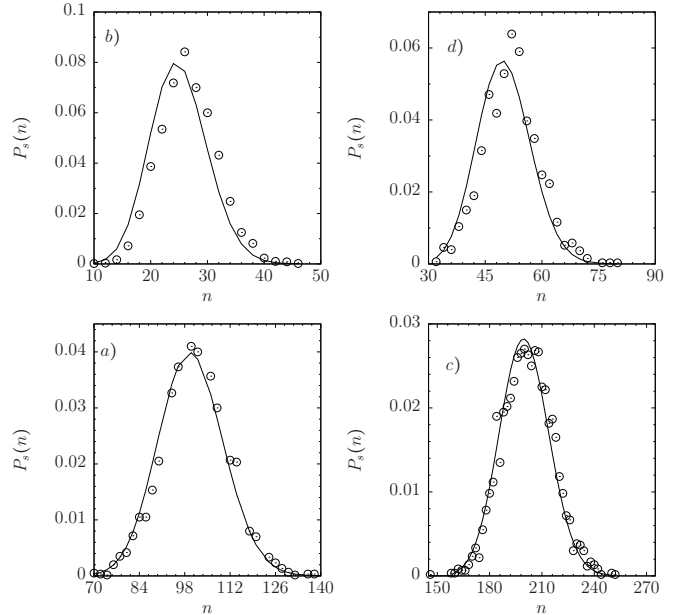}
\caption{Stationary PDF for $\text{X}\xrightarrow{\lambda}2\,
\text{X}$,
$2\,\text{X}\xrightarrow{\mu}\text{X}$. In panel a) $N=100$, in b) $N=25$, in c) $N=200$ and in d) $N=50$. Simulation results (symbols) are based
on 3000 realizations of the stochastic process up to
time $10^{6}$.}
\label{fig:f1}
\end{figure}

In this case we expect the population to evolve to
a nontrivial steady state and not to become extinct.
The equation for the probability
generating
function, (\ref{eq:G}), becomes
\begin{equation}
\frac{\partial G(p,t)}{\partial t}=\lambda p(p^{a}-1)
\frac{\partial G}{\partial p}+\frac{\mu}{2}(p-p^{2})
\frac{\partial^{2}G}{\partial p^{2}}.
\label{eq:pgf1}
\end{equation}
If initially at $t=0$ the system consists of $n_{0}$ individuals,
then $P(n,0)=\delta_{n,n_{0}}$, where $\delta$ is the Kronecker
delta, and from (\ref{eq:dG}) we find $G(p,t=0)=p^{n_{0}}$.
The boundary
conditions (BCs) are ``self-generated''. Indeed, the
equality $G(p=1,t)=1$
holds at all times, due to the conservation of probability.
Equation (\ref{eq:pgf1}) has a singular point at $p=0$. Since
$G(p,t)$
must be an analytic function at $p=0$ for all times,
we require that $G(p=0,t)=0$. This condition stems
from the fact that $G(p=0,t)=P_0(t)$,
and since the population cannot go extinct,
the probability of extinction vanishes at all times.
We are interested in the steady state.
Then (\ref{eq:pgf1})
turns into
\begin{equation}
\frac{\mu}{2}(1-p)G_{s}''+\lambda(p^{a}-1)G_{s}'=0,
\end{equation}
which must be solved with the BCs $G_{s}(1)=1$ and
$G_{s}(0)=0$. The exact analytical solution reads
\begin{equation}
G_{s}(p)=\frac{\int_{0}^{p}\exp[N\phi(s)/a]ds}{\int_{0}^{1}
\exp[N\phi(s)/a]ds},
\label{eq:Gs1}
\end{equation}
where
\begin{equation}
\phi(s)=-\ln(1-s)-\int\frac{s^{a}}{1-s}ds=\sum_{n=1}^{a}\frac{s^{n}}{n},
\label{eq:phi}
\end{equation}
and $N=2a\lambda/\mu$.
In the special case where $a=1$, the exact
solution for the generating function can be easily
obtained from (\ref{eq:Gs1})
and (\ref{eq:phi}),
\begin{equation}
G_{s}(p)=\frac{\exp(Np)-1}{\exp(N)-1}.
\label{eq:gsa1}
\end{equation}
Expanding $\exp(Np)$ around $p=0$, we find that
for large $N$ the stationary PDF follows the
Poisson distribution,
\begin{equation}
P_{s}(n)=\frac{N^{n}\exp(-N)}{n!},
\label{eq:psa1}
\end{equation}
where we have approximated $\exp(N)-1\simeq \exp(N)$
in the denominator.
We have performed numerical simulations and compared
them
with (\ref{eq:psa1}).
Figure~\ref{fig:f1} shows that the agreement becomes better
as the typical
number of individuals $N$ is increased.

\begin{figure}[htbp]
\includegraphics[width=\hsize]{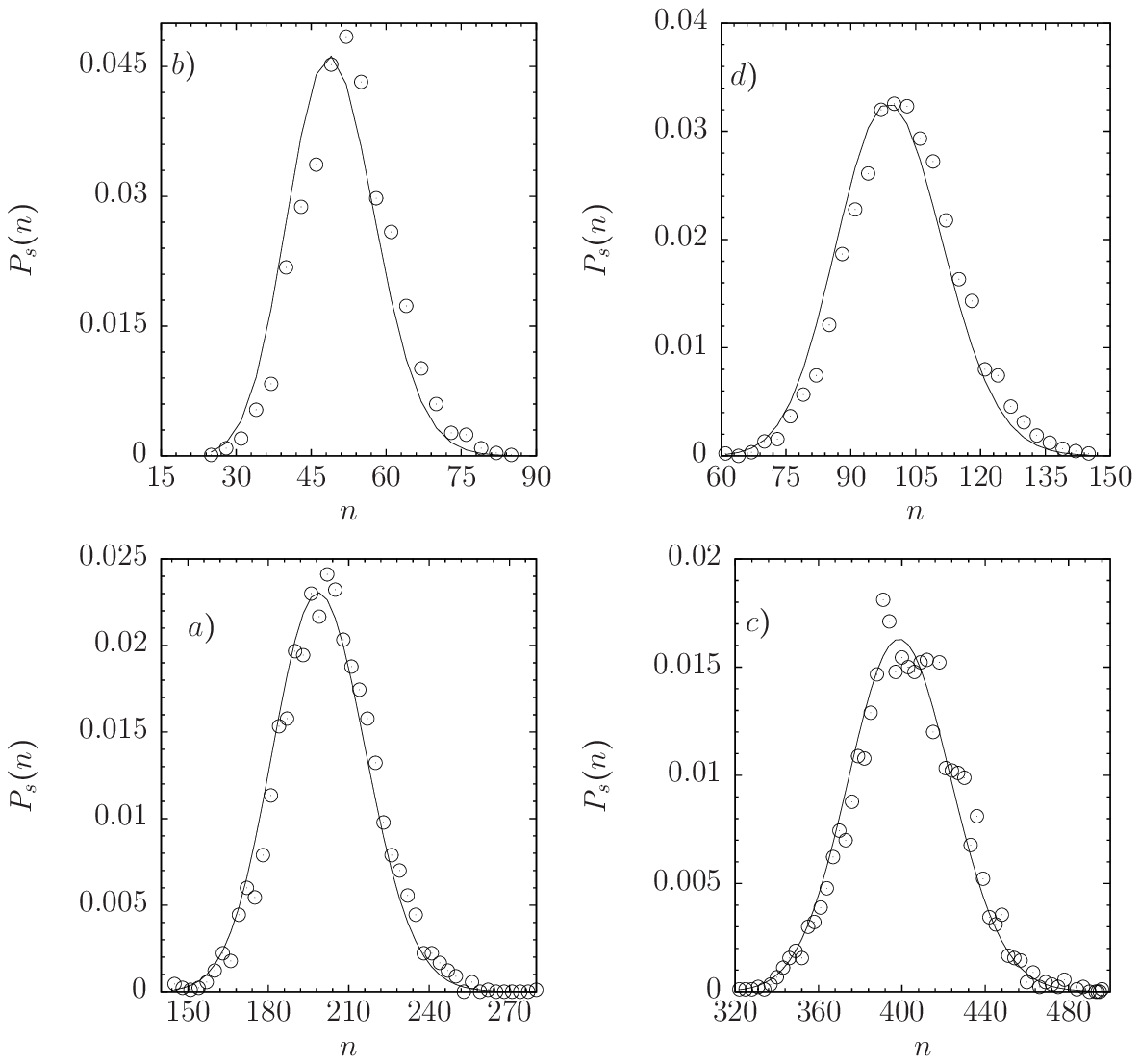}
\caption{Stationary PDF for $\text{X}\xrightarrow{\lambda}3\,
\text{X}$,
$2\,\text{X}\xrightarrow{\mu} \text{X}$.  In panel a) $N=200$, in b) $N=50$, in c) $N=400$ and in d) $N=100$. Simulation (symbols) results
are based
on 3000 realizations of the stochastic process up to
time $10^{6}$.}
\label{fig:f2}
\end{figure}

If $a=2$, the generating function is given by
\begin{equation}
G_{s}(p)=\frac{\text{erfi}\left(\frac{\sqrt{N}}{2}\right)
-\text{erfi}\left(\frac{(1+p)
\sqrt{N}}{2}\right)}{\text{erfi}\left(\frac{\sqrt{N}}{2}\right)
-\text{erfi}\left(\sqrt{N}
\right)},
\label{eq:gsa12}
\end{equation}
where $\text{erfi}(x)=\frac{2}{\sqrt{\pi}}\int_{0}^{x}
\exp(t^{2})dt$. The PDF can be
obtained by substituting (\ref{eq:gsa12})
into (\ref{Pnt}). A comparison between the
analytical PDF and numerical simulations is shown
in Fig.~\ref{fig:f2}, and excellent
agreement is observed.

\begin{figure}[htbp]
\includegraphics[width=0.8\hsize]{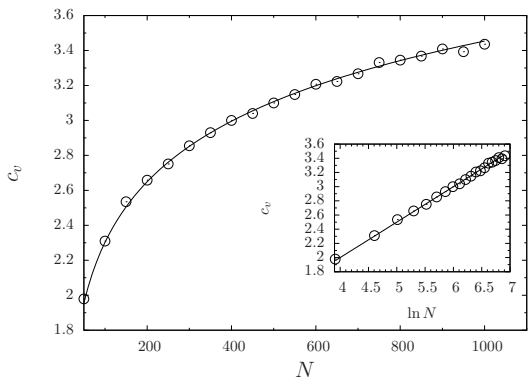}
\caption{Coefficient of variation $c_{v}$
versus $N$ for $a=1$ and $a=2$.
The log-log plot in the inset shows that
$c_{v}$ decays
like $N^{-1/2}$. Simulation results are based
on 3000 realizations of the stochastic process up to
time $10^{6}$. We set $\mu =2$ and $d=1$ and vary $\lambda$.}
\label{fig:f3}
\end{figure}

Finally, we can also obtain the mean number of
individuals in the stationary
state and its dependence on $N=2a\lambda/\mu$
by using the definition of $G$ from
(\ref{eq:dG}). Differentiating (\ref{eq:Gs1})
with respect to $p$ and using
(\ref{eq:phi}), we find
\begin{equation}
\langle n\rangle=G'(1)=\frac{\exp[\frac{N}{a}\phi(1)]}
{\int_{0}^{1}\exp[\frac{N}{a}
\phi(s)]ds}.
\label{28}
\end{equation}
Furthermore, the variance of $n$ satisfies
$\langle n^{2}\rangle-\langle n\rangle^{2}
=G''(1)+G'(1)-G'(1)^2$, and we find
\begin{equation}
\langle n^{2}\rangle-\langle n\rangle^{2}=
\langle n\rangle(1+N)-\langle n\rangle^{2}.
\label{29}
\end{equation}
This allows us to determine the
coefficient of variation, $c_{v}$, defined as
the ratio of the standard deviation to the mean,
which measures
the variability in relation to the mean
of the population,
\begin{equation}
c_{v}\equiv\frac{\sqrt{\langle n^{2}\rangle-\langle n\rangle^{2}}}
{\langle n\rangle}=
\sqrt{\frac{(1+N)}{\langle n\rangle}-1}.
\label{eq:rf}
\end{equation}
In Fig.~\ref{fig:f3} we plot the coefficient
of variation $c_{v}$ obtained from
numerical simulations (circles) and compare it
with the theoretical result
given by (\ref{eq:rf}).

It is straightforward to obtain asymptotic expressions
for the mean, the variance,
and the coefficient of variation if $N$ is large.
In that case, the integral in the denominator
of (2.9) can be evaluated by integration by parts
for Laplace integrals,
and we find
\begin{equation}
\int_{0}^{1}\exp\left[\frac{N}{a}\phi(s)\right]ds\simeq\frac{1}{N}
\left(\exp\left[\frac{N}{a}\phi(1)\right]-a\right).
\end{equation}
As a result, the mean value reads
\begin{equation}
\left\langle n\right\rangle \simeq N\left(1+a
\exp\left[-\frac{N}{a}\phi(1)\right]
+\dotsb\right),
\end{equation}
and the coefficient of variation is given by
\begin{equation}
c_{v}=\frac{1}{\sqrt{N}}\left(1-\frac{a}{2}N
\exp\left[-\frac{N}{a}\phi(1)\right]
+\dotsb\right).
\end{equation}

\subsection{Case II: $\text{\rm X}\xrightarrow{\lambda}(a+1)\,
\text{\rm X}$,
$2\,\text{\rm X}\xrightarrow{\mu}\emptyset$}

In this case, the initial number of individuals and the
parameter $a$
play a crucial role in determining the ultimate fate of
the population.
Since the death process
involves two individuals, population extinction is guaranteed,
regardless of the initial number of individuals,
if the number
of newborn individuals $a$ is odd, i.e.,
$a+1$ is even.
In contrast, if $a$ is even, i.e., $a+1$ is odd, the population becomes
extinct only if $n_{0}$
is even.

\subsubsection{$a$ is even and $n_{0}$ is odd}

\begin{figure}[htbp]
\includegraphics[width=\hsize]{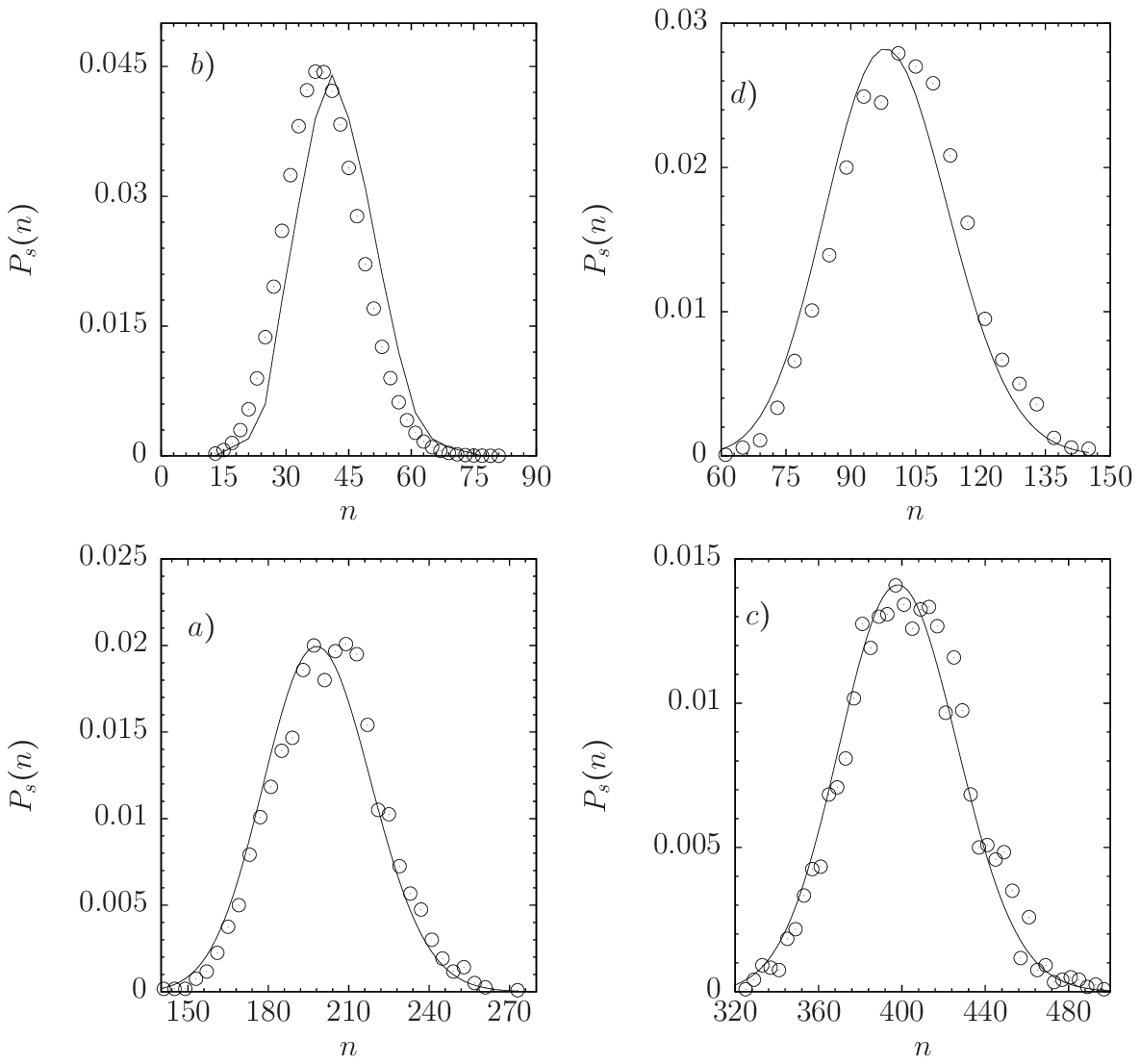}
\caption{Stationary PDF for $\text{X}\xrightarrow{\lambda}3\,\text{X}$,
$2\,\text{X}\xrightarrow{\mu}\emptyset$.  In panel a) $N=200$, in b) $N=40$, in c) $N=400$ and in d) $N=100$. Simulation (symbols) results are based
on 3000 realizations of the stochastic process up to time $10^{6}$.}
\label{fig:f4}
\end{figure}

We begin by considering the case
 that $a$ is even. Then the birth
process preserves
the even-odd parity of the number of particles. As a result, the
population becomes eventually extinct
if the initial number of individuals $n_{0}$ is even.
If $n_{0}$ is odd, the case considered in this section,
the population evolves to a nontrivial stationary
state.
The equation for the probability generating
function, (\ref{eq:G}), is given by
\begin{equation}
\frac{\partial G(p,t)}{\partial t}=\lambda p(p^{a}-1)
\frac{\partial G}{\partial p}+\frac{\mu}{2}(1-p^{2})
\frac{\partial^{2}G}{\partial p^{2}}.
\label{eq:pgf1-1}
\end{equation}
The boundary condition $G(p=1,t)=1$ still applies,
but the singular point of (\ref{eq:pgf1-1}) occurs
at $p=-1$ and not at $p=0$ as in
\eqref{eq:pgf1}.
Since
$G(p,t)$ must be analytic at $p=-1$ for all times, we require
that $G(p=-1,t)=(-1)^{n_{0}}$. This boundary condition
stems from the fact that $G(p=-1,t)$ is the sum of all
even probabilities minus the sum of
all odd probabilities \cite{AsMe06}.
The steady state has to be solved by integrating the equation
\begin{equation}
\frac{\mu}{2}(1-p^{2})G_{s}''+\lambda p(p^{a}-1)G_{s}'=0,
\end{equation}
with the boundary conditions
$G_{s}(1)=1$ and $G_{s}(-1)=(-1)^{n_{0}}$.
The exact solution reads
\begin{equation}
G_{s}(p)=C_{1}\int^{p}\exp\left[N\varphi(s)/a\right]ds+C_{2},
\end{equation}
where
\begin{equation}
\varphi(s)=-\ln(1-s^{2})-2\int\frac{s^{a+1}}{1-s^{2}}ds,
\label{eq:vfi}
\end{equation}
and $N=a\lambda/\mu$. For $n_{0}$ odd, we obtain from the
boundary conditions
\begin{equation}
C_{1}=\frac{2}{\int_{-1}^{1}\exp\left[N\varphi(s)/a\right]ds},
\end{equation}
and
\begin{equation}
C_{2}=1-\frac{2}{\int_{-1}^{1}\exp\left[N\varphi(s)/a\right]ds}.
\end{equation}
As expected,
the system reaches a nontrivial stationary state with
\begin{equation}
G_{s}(p)=1+2\frac{\int_{1}^{p}\exp\left[N\varphi(s)/a\right]ds}
{\int_{-1}^{1}\exp\left[N\varphi(s)/a\right]ds}.
\label{eq:Gs2}
\end{equation}
To be specific, we focus on the case $a=2$,
that is $\text{X}\xrightarrow{\lambda}3\,\text{X}$,
$2\,\text{X}\xrightarrow{\mu}\emptyset$.
From (\ref{eq:vfi})
we obtain $\varphi(s)=s^{2}$, and from (\ref{eq:Gs2})
\begin{equation}
G_{s}(p)=\frac{\text{erfi}\left(\frac{\sqrt{2N}p}{2}\right)}
{\text{erfi}\left(\frac{\sqrt{2N}}{2}\right)}.\label{eq:Gs3}
\end{equation}
The PDF is obtained by substituting (\ref{eq:Gs3}) into (\ref{Pnt}).
In Fig.~\ref{fig:f4} we plot the PDF $P_{s}(n)$ for different values
of $N$.

\begin{figure}[htbp]
\includegraphics[width=0.9\hsize]{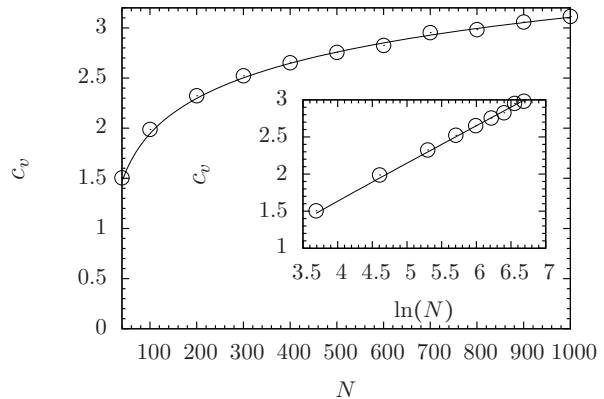}
\caption{Coefficient of variation versus $N$ for $a=2$ and $a=4$.
The inset demonstrates that $c_{v}$ decays
like $N^{-1/2}$. Simulation results
are based on
3000 realizations of the stochastic process
up to time $10^{6}$. We set $\mu=d=2$ and vary $\lambda$.}
\label{fig:f5}
\end{figure}

In Fig.~\ref{fig:f5} we plot the coefficient of variation
$c_{v}$ for the cases of $a=2$ and $a=4$.
The mean number of individuals in the steady state,
$\langle n\rangle=G'(1)$, can be determined
from (\ref{eq:Gs3}),
\begin{equation}
\langle n\rangle=\frac{\sqrt{2N}\exp(N/2)}{\sqrt{\pi}\,
\text{erfi}(\sqrt{2N}/2)},
\label{nm2}
\end{equation}
and the coefficient of variation is given by (\ref{eq:rf}) with
$\langle n\rangle$ given by (\ref{nm2}). The solid curve corresponds
to the analytical results, and the symbols correspond to numerical
simulations. The inset again shows that $c_{v}$ scales like $N^{-1/2}$.

\subsubsection{$a$ and $n_{0}$ are even}

If $n_{0}$ is even, the boundary conditions
lead to
\begin{equation}
C_{1}\int^{1}\exp[N\varphi(s)/a]ds+C_{2}=1,
\label{eq:C1aeven}
\end{equation}
and
\begin{equation}
C_{1}\int^{-1}\exp[N\varphi(s)/a]ds+C_{2}=1,
\label{eq:C2aeven}
\end{equation}
so that $C_{1}=0$
and $C_{2}=1$. Therefore, $G_{s}(p)=1$ which describes an empty
population state, i.e., extinction, as $t\to\infty$.
To calculate
the MET, we employ the ``momentum-space" spectral method developed
recently \cite{KaEl04,AsMe06,AsMe07,AsMeSa10}.
After a short relaxation time $t_r$, which corresponds
to the deterministic relaxation time of the system
to the stable stationary state, the population typically
settles into a long-lived metastable state,
which is encoded by the lowest
excited eigenmode $\psi(p)$ of the probability generating
function $G(p,t)$ \cite{AsMeSa10}. Indeed,
for $t\gg t_r$, we can write
\begin{equation}
G(p,t)=G_{s}(p)-\psi(p)\exp\left(-\mu E_{1}t\right).
\label{eq:gap}
\end{equation}
Here $E_{1}$ is the lowest nonzero eigenvalue,
$\tau=(\mu E_{1})^{-1}$ is the
mean time to extinction, and $G_{s}(p)=1$.
Substituting (\ref{eq:gap})
into (\ref{eq:pgf1}), we obtain
\begin{equation}
(1-p^{2})\psi''(p)+2\Omega p(p^{a}-1)\psi'(p)=-2E_{1}\psi(p),
\label{eq:efpsi}
\end{equation}
where $\Omega\equiv\lambda/\mu$. Since $a$ is even,
the function $\psi(p)$
is also an even function. It is therefore
sufficient to consider the interval
$0\leq p<1$. Since $\langle n\rangle\sim \Omega$,
we assume that $\Omega\gg 1$ to find the eigenvalue $E_{1}$,
which we expect to be exponentially small in $\Omega$.
We will proceed by matching the asymptotic expansion
for the function $\psi(p)$
in the bulk region, $0\leq p<1$, namely $\psi_{b}$,
with $\psi_{l}$, the solution in the boundary layer, $1-p\ll 1$.
We will show that the function $\psi(p)$ is almost
constant everywhere within the
interval $p\in[0,1)$, except in a narrow layer close to $p=1$.
In the bulk we can treat $E_{1}$ as a perturbative parameter. To
zero order we set $E_{1}=0$,
and the even solution of (\ref{eq:efpsi})
is 1. To account for corrections, we write
$\psi(p)=1+\delta\psi$, where $\delta\psi\ll 1$
satisfies the differential equation
\begin{equation}
\delta\psi''+2\Omega p\frac{p^{a}-1}{1-p^{2}}
\delta\psi'=-\frac{2E_{1}}{1-p^{2}},
\label{eq:dfi}
\end{equation}
whose solution, using Eq.~(\ref{eq:vfi}), takes the form
\begin{multline}
\delta\psi'(p)=C_{0}\exp\left[\Omega\varphi(p)\right]\\
-2E_{1}
\exp\left[\Omega\varphi(p)\right]
\int^{p}\frac{\exp\left[-\Omega\varphi(s)\right]}{1-s^{2}}ds.
\label{eq:soldf}
\end{multline}
To solve for $\psi(p)$, we need to specify two boundary conditions.
Setting $p=0$ in (\ref{eq:efpsi}), we obtain $\psi''(0)=-2E_{1}\psi(0)$,
or equivalently $\delta\psi''(0)=-2E_{1}-2E_{1}\delta\psi(0)$. On
the other hand, from (\ref{eq:dfi}) and setting $p=0$, we find the
first boundary condition, $\delta\psi''(0)=-2E_{1}$. This condition
together with
$\delta\psi''(0)=-2E_{1}-2E_{1}\delta\psi(0)$ leads
to the second boundary condition, $\delta\psi(0)=0$.
The
first boundary condition implies
that (\ref{eq:soldf}) reduces to
\begin{equation}
\delta\psi'(p)=-2E_{1}\exp\left[\Omega\varphi(p)\right]
\int_{0}^{p}\frac{\exp\left[-\Omega\varphi(s)\right]}{1-s^{2}}ds,
\label{eq:soldf1}
\end{equation}
which can be integrated together with the second boundary condition
to yield
\begin{equation}
\delta\psi(p)=-2E_{1}\int_{0}^{p}\exp\left[\Omega\varphi(s)\right]ds
\int_{0}^{s}\frac{\exp\left[-\Omega\varphi(u)\right]}{1-u^{2}}du.
\label{eq:dff}
\end{equation}
Since this solution holds in the bulk region $1-p\gg\Omega^{-1}$, with
$\Omega\gg1$, we can approximate the inner integral in (\ref{eq:dff})
as follows
\begin{multline}
\int_{0}^{s}\frac{\exp\left[-\Omega\varphi(u)\right]}{1-u^{2}}du
\simeq
\int_{0}^{s}\exp\left[-\Omega\varphi(u)\right]du\\
\simeq\int_{0}^{\infty}\!\!\exp\left[-\Omega\varphi(u)\right]du.
\label{eq:ap1}
\end{multline}
Therefore,
\begin{equation}
\psi_{b}(p)\simeq 1-2E_{1}\int_{0}^{p}\exp
\left[\Omega\varphi(s)\right]ds\int_{0}^{\infty}\!\!
\exp\left[-\Omega\varphi(u)\right]du.
\label{eq:fb}
\end{equation}
In the boundary layer, $1-p\ll 1$, we disregard the exponentially small
term $E_{1}\psi$ in (\ref{eq:efpsi}) and integrate the resulting
equation $(1-p^{2})\psi''(p)+2\Omega p(p^{a}-1)\psi'(p)=0$ to obtain
\begin{equation}
\psi_{l}(p)=C\int_{1}^{p}\exp\left[\Omega\varphi(s)\right]ds,
\label{eq:fl}
\end{equation}
where we have made use of the boundary condition
at the boundary layer, i.e., $\psi_{l}(1)=0$.
Equation (\ref{eq:fl}) can be rewritten as
\begin{multline}
\psi_{l}(p)=C\left(\int_{0}^{p}\exp\left[\Omega\varphi(s)\right]ds
-\int_{0}^{1}\exp\left[\Omega\varphi(s)\right]ds\right)\\
=C_{1}\left(1-\frac{\int_{0}^{p}\exp\left[\Omega\varphi(s)\right]ds}
{\int_{0}^{1}\exp\left[\Omega\varphi(s)\right]ds}\right).
\label{eq:fl2}
\end{multline}
Matching the solutions (\ref{eq:fb}) and (\ref{eq:fl2}),
we find
$C_{1}=1$ and the MET,
\begin{equation}
\tau=\frac{2}{\mu}\int_{0}^{1}\exp\left[\Omega\varphi(s)\right]ds
\int_{0}^{\infty}\exp\left[-\Omega\varphi(u)\right]du.
\label{eq:tau1}
\end{equation}
Since $\Omega\gg1$, we can further approximate (\ref{eq:tau1}). The
function $\varphi(s)$, given by (\ref{eq:vfi}),
can be expressed
as
\begin{equation}
\varphi(s)=\sum_{j=1}^{a/2}\frac{s^{2j}}{j}
\label{eq:fs1}
\end{equation}
for even $a$, and
\begin{equation}
\varphi(s)=-2\ln(1+s)+2\sum_{j=0}^{(a-1)/2}\frac{s^{2j+1}}{2j+1}
\label{eq:fs2}
\end{equation}
for odd $a$. Since in this subsection we consider the case of even $a$,
$\varphi(s)$ is a polynomial of order $a$ with positive coefficients.
Therefore, the main contribution of the first integral in
(\ref{eq:tau1}) comes from the region
around $s=1$. Employing the Taylor expansion we find
\begin{eqnarray}
&&\hspace{-5mm}\int_{0}^{1}\exp\left[\Omega\varphi(s)\right]ds
\simeq\int_{0}^{1}
\exp\left\{\Omega\left[\varphi(1)+\varphi'(1)(s-1)\right]\right\}ds\nonumber\\
&&\simeq\frac{\exp\left[\Omega\varphi(1)\right]}{\Omega\varphi'(1)}.
\end{eqnarray}
For the second integral in (\ref{eq:tau1}),
the main contribution comes from the region
around $u=0$. To leading order, $\varphi(u)\simeq u^{2}$ and
\begin{equation}
\int_{0}^{\infty}\exp\left[-\Omega\varphi(u)\right]du
\simeq\int_{0}^{\infty}\exp\left[-\Omega u^{2}\right]du=
\frac{\sqrt{\pi}}{2\sqrt{\Omega}}.
\label{eq:ap2}
\end{equation}
Substituting these results into (\ref{eq:tau1}), we
obtain a general result for the MET for $\Omega\gg 1$
and any even $a$,
\begin{equation}
\tau=\frac{\sqrt{\pi}\exp\left(\Omega\sum_{j=1}^{a/2}
\frac{1}{j}\right)}{\mu a\Omega^{3/2}}.\label{eq:tauf}
\end{equation}

\begin{figure}[htbp]
\includegraphics[width=0.9\hsize]{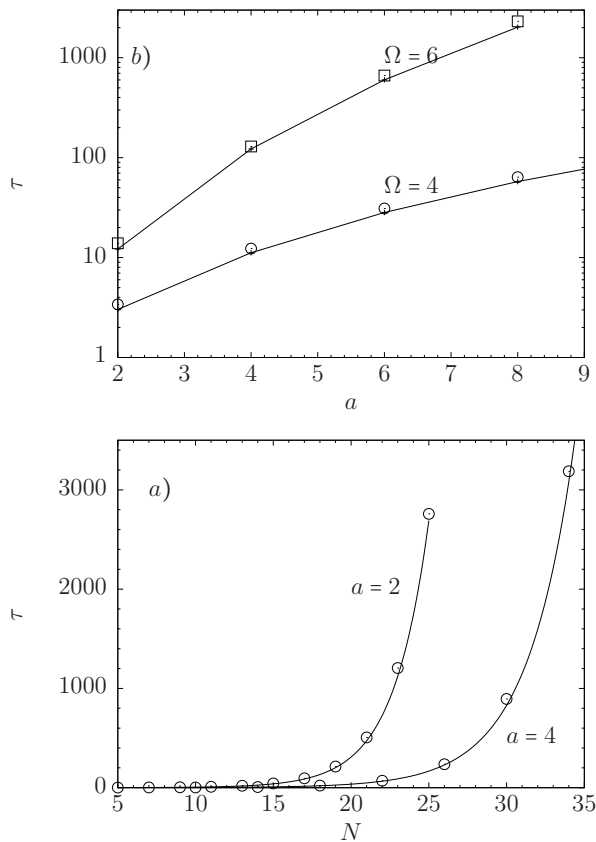}
\caption{Mean time to extinction $\tau$ vs $N$ (panel a))
and vs $a$ (panel b)) for reaction
$\text{X}\xrightarrow{\lambda}(a+1)\,\text{X}$,
$2\,\text{X}\xrightarrow{\mu}\emptyset$.  Solid curves are obtained from
(\ref{eq:tauf}), while symbols correspond to numerical simulations.
We set $\mu =2$ and vary $\lambda$.
Simulations have been performed up to time $10^{8}$.}
\label{fig:f67}
\end{figure}

As an example, for $a=2$ we find
\begin{equation}
\tau=\frac{\sqrt{\pi}\exp\left(\Omega\right)}{2\mu \Omega^{3/2}},
\end{equation}
which coincides with the result in \cite{AsMe06}.
We have verified the result (\ref{eq:tauf}) by
numerical simulations.  In the upper panel
of Fig.~\ref{fig:f67},
we plot $\tau$ versus
$N$ for $a=2$ and $a=4$, and
in the lower panel we plot $\tau$ versus $a$
for different values of $\Omega$.
In all these comparisons we obtain excellent
agreement between theory and simulations.

\subsubsection{$a$ is odd}

If $a$ is odd, (\ref{eq:pgf1-1}) has no other singularity and
we have only one boundary condition, $G_{s}(1)=1$.
As a result, $G_{s}(p)=1$,
and the population becomes extinct, regardless
of the value of $n_{0}$.
To obtain the MET in this case, we start again
with (\ref{eq:pgf1-1}).
Since $\psi(p)$ is no longer even, the bulk region now
corresponds to $p\in[-1,1)$,
and the boundary layer is located at $1-p\ll 1$. In the bulk
region we impose the boundary condition
$\delta\psi(0)=0$, as in the case of even $a$. However, setting
$p=-1$ in (\ref{eq:efpsi}), we find now the second boundary
condition to be $\psi'(-1)=0$, where we
have neglected the term $E_{1}\psi(-1)$,
 which is exponentially small.
The final solution for the function $\psi$ in the bulk region is very
similar to the even $a$ case, and we find
\begin{equation}
\psi_{b}(p)=1-2E_{1}\int_{0}^{p}\exp\left[\Omega\varphi(s)\right]ds
\int_{-1}^{s}\frac{\exp\left[-\Omega\varphi(u)\right]}{1-u^{2}}du.
\label{eq:dff2}
\end{equation}
In the boundary layer we obtain exactly the same result as
(\ref{eq:fl}).
By matching both solutions in the common region,
we obtain
\begin{equation}
\tau=\frac{2}{\mu}\int_{0}^{1}\exp\left[\Omega\varphi(s)\right]ds
\int_{-1}^{\infty}\frac{\exp\left[-\Omega\varphi(u)\right]}
{1-u^{2}}du.
\label{eq:tau2}
\end{equation}
To proceed, we employ the approximations (\ref{eq:ap1})
and
\begin{multline}
\int_{-1}^{s}\frac{\exp\left[-\Omega\varphi(u)\right]}{1-u^{2}}du
\simeq\int_{-1}^{s}
\exp\left[-\Omega\varphi(u)\right]du\\
\simeq\int_{-1}^{\infty}\exp\left[-\Omega\varphi(u)\right]du
\simeq\int_{-1}^{\infty}\exp\left[-\Omega u^{2}\right]du=
\frac{\sqrt{\pi}}{\sqrt{\Omega}}.
\end{multline}
As a result, similar to the even $a$ case,
we obtain from (\ref{eq:fs2})
the general result for any odd $a$,
\begin{equation}
\tau=\frac{2\sqrt{\pi}}{\mu a\Omega^{3/2}}\exp
\left( -2\Omega\ln2+2\Omega\sum_{j=0}^{\frac{a-1}{2}}
\frac{1}{2j+1}\right).
\label{tau21}
\end{equation}
For $a=1$, (\ref{tau21}) yields
\begin{equation}
\tau=\frac{2\sqrt{\pi}\exp\left[2\Omega(1-\ln2)\right]}
{\mu\Omega^{3/2}}
\end{equation}
which coincides with the result in \cite{AsMe07}.

\begin{figure}[htbp]
\includegraphics[width=0.9\hsize]{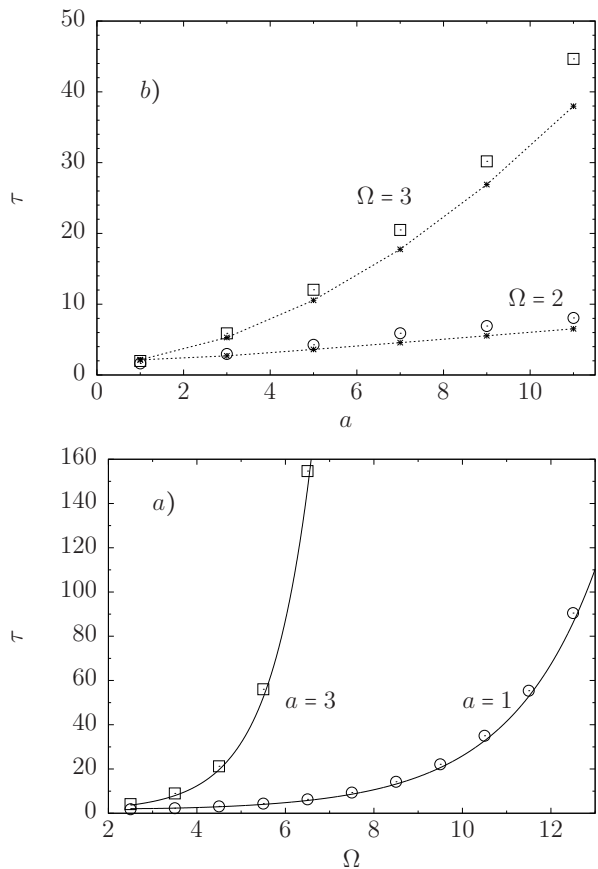}
\caption{Mean time to extinction $\tau$ vs $\Omega$ (panel a))
and vs $a$ (panel b)) for reaction $\text{X}
\xrightarrow{\lambda}(a+1)\,\text{X}$,
$2\,\text{X}\xrightarrow{\mu}\emptyset$. Solid curves are obtained from
(\ref{tau21}), while symbols correspond to numerical simulations.
We set $\mu =2$
and vary $\lambda$.
Simulations have been performed up to time $10^{8}$.}
\label{fig:f89}
\end{figure}

In Fig.~\ref{fig:f89} we verify the result (\ref{tau21})
for the MET. In the upper panel we plot $\tau$ versus $N$
for $a=1$ and $a=3$. The mean time to extinction
increases as the number of individuals increases, as expected.
In the lower panel we plot $\tau$ versus $a$ for
relatively low values of $\Omega$,
and the agreement between theory and numerical simulations
is still fair.

\section{Birth-competition-death processes}

We add the death reaction $\text{X}\xrightarrow{\gamma}\emptyset$
to the system of birth-competition processes (\ref{eq:bd}). To
obtain a logistic equation in the mean-field limit, we consider
$b=1$ and $c=2$, leading to the reaction scheme
\begin{subequations}
\label{3e}
\begin{align}
\text{X} & \xrightarrow{\lambda}  (a+1)\,\text{X},
\label{eq:3e1}\\
2\,\text{X} & \xrightarrow{\mu} (2-d)\,\text{X},
\label{eq:3e2}\\
\text{X} & \xrightarrow{\gamma} \emptyset.
\label{eq:3e3}
\end{align}
\end{subequations}
Here, $a\geq 1$ and $d=1$ for a birth-competition-death system and
$d=2$ for a birth-annihilation-death system. It is
straightforward to show that
this system always goes extinct. We are interested in calculating
the MET for the general case. Although this can also be
done via the generating function ($p$-space theory),
we will use the ``real-space" WKB approximation \citep{kessler2007extinction,meerson2008noise,escudero2009switching,AsMe10}.
According to \eqref{trans}, the
transition rates are given by
\begin{subequations}
\label{rates}
\begin{align}
W(n,a)&=\lambda n,\\
W(n,-d)&=\frac{\mu}{2}\frac{n!}{(n-2)!}=\frac{\mu}{2}n(n-1),\\
W(n,-1)&=\gamma n.
\end{align}
\end{subequations}
Replacing $t$ by $t/\gamma$ and introducing the rescaled population
number density $q=n/N$, where $N=\lambda/\mu\gg1$,
the transition rates can be
rewritten as
\begin{equation}
W(n,r)\equiv W(Nq,r)=Nw_{r}(q)+u_{r}(q)+O(N^{-1}),
\end{equation}
where
\begin{subequations}
\begin{align}
w_{a}(q)&=R_{0}q,\\
w_{-d}(q)&=\frac{1}{2}R_{0}q^{2},\\
w_{-1}(q)&=q.
\end{align}
\end{subequations}
Here
$q$, $w_r(q)$, and $u_r(q)$ are $O(1)$, and
\begin{subequations}
\begin{align}
u_{a}(q)&=u_{-1}(q)=0,\\
u_{-d}(q)&=-\frac{1}{2}R_{0}q.
\end{align}
\end{subequations}
Further, $R_{0}=\lambda/\gamma$ is the basic reproductive number.
Since $n=q=0$ is an absorbing
state (extinction), we have $w_{r}(0)=u_{r}(0)=0$
for any $r=\{a,-d,-1\}$.
For $N\gg 1$,
the WKB theory developed in \cite{kessler2007extinction,meerson2008noise,escudero2009switching,AsMe10} can be used
for the rescaled master equation. Accordingly, we look for
the probability $P(n,t)=P(Nq,t)$ in the form of the
WKB ansatz
\begin{equation}
P(q,t)=\exp\left[-NS(q)\right]
\label{wkb}
\end{equation}
where $S(q)$ is a deterministic state function known as
the action.
Intuitively, this approximation expresses the assumption
that the probability of occurrence of extreme events,
such as extinction, lies in the tail of the PDF, which falls
away steeply from the steady state.
Substituting (\ref{wkb}) into
the rescaled master equation (\ref{eq:me2}),
which contains terms
of the form $w_r(q-r/N)$,
and Taylor-expanding terms such as $S(q-r/N)$ around $q$,
we obtain to leading order a Hamilton-Jacobi equation
$H(p,q)=0$ \cite{Dy94}, with Hamiltonian
\begin{multline}
H(p,q)=\sum_{r}w_{r}(q)\left[\exp(rp)-1\right]\\
=R_{0}q[\exp(ap)-1]+\frac{R_{0}}{2}q^{2}[\exp(-dp)-1]+q[\exp(-p)-1].
\label{H}
\end{multline}
Here $q$ is the coordinate, and $p=S'(q)$ is the conjugate momentum.
The mean-field dynamics can be found by writing the Hamilton's
equation $\dot{q}=\partial_p H$ along
the path $p=0$. This yields the logistic equation as
the mean-field description of the system (\ref{3e}),
\begin{eqnarray}
&&\frac{d\langle q\rangle}{dt}=\left(\frac{\partial H}{\partial p}\right)_{p=0}=
\sum_{r}rw_{r}(\langle q\rangle)\nonumber\\
&&=\langle q\rangle\left(aR_{0}-1-\frac{dR_{0}}{2}\langle q\rangle\right).
\label{MF3e}
\end{eqnarray}
Equation \eqref{MF3e} has an nontrivial attracting steady state
at
\begin{equation}
q_{*}=\frac{2}{d}\left(a-1/R_{0}\right),
\end{equation}
if
\begin{equation}
aR_{0}>1.
\label{cc}
\end{equation}
Note that a bifurcation occurs at $R_0=1/a$. This implies that
the population can maintain a long-lived metastable state for $a>1$,
even if $R_0<1$. Going back to the mean number of individuals $n$,
the logistic mean-field rate equation (\ref{MF3e}) reads
\begin{equation}
\frac{dn}{dt}=rn\left( 1-\frac{n}{K}\right),
\end{equation}
where
\begin{equation}
r\equiv aR_0-1=\frac{a\lambda-\gamma}{\mu}
\label{r}
\end{equation}
and
\begin{equation}
K\equiv Nq_*=\frac{2(a\lambda-\gamma)}{d\mu}
\label{K}
\end{equation}
are the intrinsic growth rate and the carrying capacity, respectively.
The mean-field logistic equation is completely specified
by the two parameters $r$ and $K$.
On the other hand, the stochastic dynamics is characterized
by up to five parameters, namely
$a$, $d$, $\lambda$, $\mu$ and
$\gamma$. Relations (\ref{r}) and
(\ref{K}) provide a
constraint for two of them,
and the other three are free to take different values while
keeping exactly the same mean-field logistic equation.

In order to find the MET, we need to find the optimal path to
extinction, which is defined as the nontrivial heteroclinic orbit
that solves the equation $H(q,p)=0$ in the phase
space $(q,p)$ and connects the state $(q_{*},p=0)$ to the extinction
state $(q=0,p=p_{f})$. Here $p_{f}$ is the
solution to the equation
$q_{a}(p_{f})=0$, namely, the value of the momentum along
the optimal path to extinction at the point where $q$ vanishes.
For the system (\ref{3e}) we find the optimal path to extinction
(activation trajectory)
\begin{equation}
q_{a}(p)=2\frac{R_{0}\left[\exp(ap)-1\right]-1+\exp(-p)}{R_{0}
\left[1-\exp(-dp)\right]},
\label{qa}
\end{equation}
and $p_{f}$ is the solution of the transcendental equation
\begin{equation}
R_{0}\left[\exp(ap_{f})-1\right]-1+\exp(-p_{f})=0.
\label{eqtr}
\end{equation}
According to \cite{AsMe10}, the MET is given by
\begin{equation}
\tau=\frac{A_{1}\sqrt{2\pi}}{\gamma q_{*}}
\sqrt{\frac{q_{a}'(p=0)}{N}}
\exp(N\Delta S)\exp(\Delta\phi),
\label{tausc}
\end{equation}
where, taking into account (\ref{qa}),
\begin{equation}
\frac{1}{q_{*}}
\sqrt{\frac{q_{a}'(p=0)}{N}}=
\frac{\sqrt{dR_{0}}}{2(aR_{0}-1)}
\sqrt{\frac{R_{0}a(a+d)+1-d}{N}}.
\end{equation}
The quantities $\Delta S$ and $\Delta\phi$ can be
calculated as follows.
$\Delta S$ is the action increment along the extinction path, which
gives us the logarithm of the mean time to extinction~\cite{AsMe10}.
Since $p=dS/dq$,
\begin{equation}
\Delta S=S(0)-S(q_{*})=\int_{q_{*}}^{0}p_{a}(q)dq=
\int_{p_{f}}^{0}q_{a}(p)dp.
\label{dSe3}
\end{equation}
Making use of (\ref{qa}), we obtain from (\ref{dSe3}):
\begin{multline}
\Delta S=S(0)-S(q_{*})\\
=2\int_{e^{p_{f}}}^{1}\frac{z^{a+d}
-(1+R_{0}^{-1})z^{d}+R_{0}^{-1}z^{d-1}}{z(z^{d}-1)}dz.
\label{dSg}
\end{multline}
For $d=1$, this equation yields
\begin{equation}
\Delta S=S(0)-S(q_{*})=\frac{2p_{f}}{R_{0}}+2\sum_{j=1}^{a}
\frac{1}{j}-2\sum_{j=1}^{a}\frac{\exp(jp_{f})}{j}.
\label{dS1}
\end{equation}
For $d=2$, \eqref{dSg} yields for even $a$,
\begin{multline}
\Delta S=S(0)-S(q_{*})=2\left(1+\frac{1}{R_{0}}\right)
\ln\left(\frac{1+\exp(p_{f})}{2}\right)\\
+2\sum_{j=1}^{a/2}\frac{1-\exp(2jp_{f})}{2j},
\label{dS21}
\end{multline}
and for odd $a$,
\begin{multline}
\Delta S=S(0)-S(q_{*})=2\left(1+\frac{1}{R_{0}}\right)
\ln\left(\frac{1+\exp(p_{f})}{2}\right)\\
+2\sum_{j=1}^{(a+1)/2}\frac{1-\exp[(2j-1)p_{f}]}{2j-1}.
\label{dS22}
\end{multline}

In order to go beyond leading-order calculations,
we determine $\Delta\phi=\phi(q=0)-\phi(q=q_{*})$,
using its definition given in \cite{AsMe10},
\begin{widetext}
\begin{equation}
\Delta\phi=\int_{0}^{p_{f}}q_{a}'(p)\left[\frac{H_{pq}(q_{a},p)
+\frac{1}{2}
[q_{a}'(p)]^{-1}H_{pp}(q_{a},p)+\frac{R_{0}}{2}q_{a}(p)
\left[\exp(-dp)-1\right]}
{H_{p}(q_{a},p)}-\frac{1}{q_{a}(p)}\right]dp,
\label{eq:dfi-1}
\end{equation}
where $q_{a}'(p)=dq_{a}/dp$, and the subscripts on $H$ indicate partial
derivatives. Making use of (\ref{H}) and (\ref{qa}), we obtain
from (\ref{eq:dfi-1})
for $d=1$,
\begin{equation}
\Delta\phi=-\frac{p_{f}}{2}-\frac{1}{2}
\ln\left(\frac{1+a}{2}\right)+\frac{1}{2}
\ln\left[\frac{a\exp[(a+1)p_{f}]-(1+a)\exp(ap_{f})+1}
{a\left[\exp(p_{f})-1\right]^{2}}\right],
\label{f1}
\end{equation}
and for $d=2$,
\begin{equation}
\Delta\phi=-\frac{p_{f}}{2}+\frac{1}{2}
\ln\left[4\frac{aR_{0}\exp[p_{f}(a+3)]-R_{0}
(a+2)\exp[p_{f}(a+1)]-\exp(2p_{f})+2(R_{0}+1)\exp(p_{f})-1}{(a^{2}
R_{0}+2aR_{0}-1)\left[\exp(2p_{f})-1\right]^{2}}\right].
\label{f2}
\end{equation}
\end{widetext}

The formula for the coefficient $A_1$ is given by, see
Eq.~(39) of \cite{AsMe10},
\begin{equation}
\label{A1}
A_1=\frac{(-1)^a \prod_{j=1}^a l_j}
{(l_1-1)\prod_{j=2}^a (l_1-l_j)},
\end{equation}
where $l_i$ are the roots of the equation
\begin{equation}\label{root}
w_a^{\prime}(0) l^{a+1} - \left[1+w_a^{\prime}(0)\right]l +1 =0.
\end{equation}
Here we have used Eq.~(31) of \cite{AsMe10}.
It can be shown that one root
of this equation is always $l=1$. We denote
this root by $l_0$.
Using the fact that $w_a^{\prime}(0)=R_0$, and dividing by $l-1$,
we need to solve the equation $l R_0 (1+l+\dotsb+l^{a-1})-1=0$,
i.e.,
\begin{equation}\label{root1}
l+\dotsb+l^a=\frac{1}{R_0}.
\end{equation}
For $a=1$, (\ref{root1}) has a single root, and
therefore (\ref{A1})
simplifies to $A_1^{(a=1)}=1/(R_0-1)$.
For $a=2$, (\ref{root1})
has two roots, and we find
\begin{equation}
A_1^{(a=2)}=\frac{2}{3 \sqrt{R_0^2+4R_0}-R_0-4}.
\end{equation}
For $a=3$, the polynomial of (\ref{root1}) is of third order,
and its solution yields
\begin{equation}
A_1^{(a=3)}=\frac{(c+2)(4-c)(c^2-2c+4)(c^2+4c+16)}
{3(c^2-8c-8)(c^4-8c^2+64)},
\end{equation}
where
\begin{equation}
c=\left[4\frac{3\sqrt{3}\sqrt{3R_0^2+14R_0+27}+7R_0+27}
{R_0}\right]^{1/3}.
\end{equation}
Note that the value of $A_1^{(a)}$ does not depend on $d$.
The origin of this behavior lies in the fact
that reactions without a linear term in $n$, such as
(\ref{eq:3e2}),
do not play a role in the recursive solution of the master equation
for small values of $n$ \cite{AsMe10}.

Exact analytic expressions for the MET can be obtained
for some specific cases.
For example, if $a=d=1$ (birth-competition-death),
the reactions are given by
\begin{subequations}
\label{gsys2}
\begin{align}
\text{X} & \xrightarrow{\lambda}  2\,\text{X},
\\
2\,\text{X} & \xrightarrow{\mu} \text{X},
\\
\text{X} & \xrightarrow{\gamma} \emptyset.
\end{align}
\end{subequations}
Equation (\ref{tausc}) yields for the MET,
\begin{equation}
\tau=\frac{1}{\gamma}\sqrt{\frac{\pi}{N}}\frac{R_{0}^{3/2}}
{(R_{0}-1)^{2}}\exp\left[2N\left(1-\frac{1+%
\ln R_{0}}{R_{0}}\right)\right],
\label{ta1d1}
\end{equation}
recovering the result given by Eq.~(70) in \cite{AsMe10}.

\begin{figure}[!htbp]
\includegraphics[width=0.9\hsize]{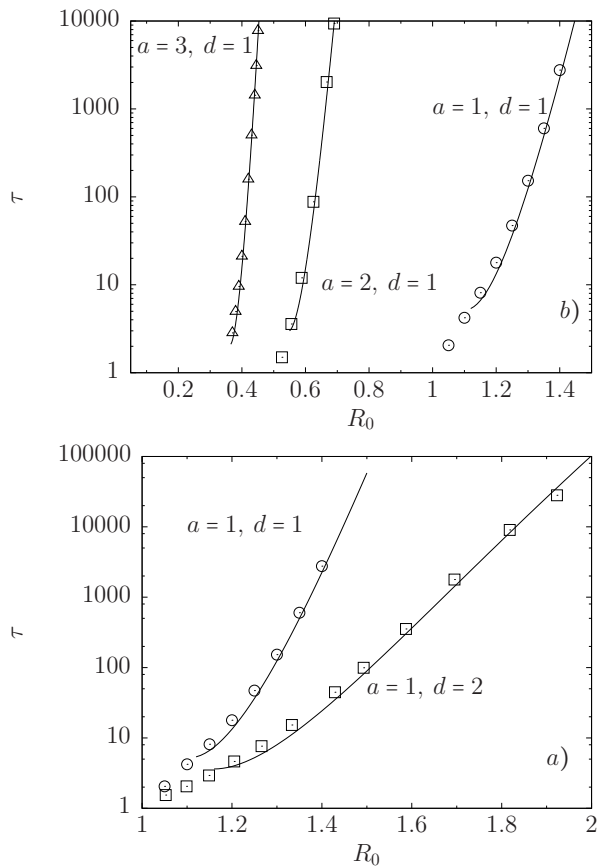} 
\caption{Mean time to extinction for birth-and-death reactions.
In both cases we set
$\mu =0.1$, $\lambda=10$, $d=1$, $N=100$ and modify $\gamma$ to vary $R_0$. In panel a) we compare different values for the MET for different $d$ and the same $a$. In panel b) we compare different values for the MET for different $a$ and the same $b$. Simulations (symbols)
have been performed up to time $10^{9}$, and mean values
are obtained by averaging over $4\times 10^4$ realizations.
Solid curves correspond to exact analytic results given by
(\ref{ta1d1}), (\ref{ta2d1}) and (\ref{ta1d2})}
\label{fig:f1011}
\end{figure}

For $a=2$ and $d=1$,
the reactions are again of birth-competition-death type,
\begin{subequations}
\label{gsys3}
\begin{align}
\text{X} & \xrightarrow{\lambda}  3\,\text{X},
\\
2\,\text{X} & \xrightarrow{\mu} \text{X},
\\
\text{X} & \xrightarrow{\gamma} \emptyset.
\end{align}
\end{subequations}
\begin{widetext}
In this case, (\ref{tausc}) yields for the MET,
\begin{equation}
\tau=\frac{1}{\gamma}\sqrt{\frac{\pi}{N}}\frac{R_{0}
\left(3\sqrt{R_{0}(R_{0}+4)}
+R_{0}+4\right)\sqrt{R_{0}+4+\sqrt{R_{0}(R_{0}+4)}}}
{2(2R_{0}-1)^{2}(R_{0}+4)}
\exp(2N\Delta S),
\label{ta2d1}
\end{equation}
where
\begin{equation}
\Delta S=\frac{1}{R_{0}}\ln\left(\frac{2}{R_{0}
+\sqrt{R_{0}(R_{0}+4)}}\right)+
\frac{(3R_{0}-2)\sqrt{R_{0}(R_{0}+4)}+3R_{0}^{2}
+4R_{0}-2}{\left(R_{0}
+\sqrt{R_{0}(R_{0}+4)}\right)^{2}}.
\end{equation}
The difference between results (\ref{ta1d1}) and (\ref{ta2d1}) is
due the value of $a$. Figure~\ref{fig:f1011} shows
that increasing $a$  by one unit
increases the MET by
several orders of magnitude.
One can show that $\tau(a=2)/\tau(a=1)\sim \exp(N)$
as $R_0$ tends to infinity.

Finally, we consider the case $a=1$ and $d=2$,
\begin{subequations}
\label{gsys4}
\begin{align}
\text{X} & \xrightarrow{\lambda} 2\,\text{X},
\\
2\,\text{X} & \xrightarrow{\mu} \emptyset,
\\
\text{X} & \xrightarrow{\gamma} \emptyset.
\end{align}
\end{subequations}
In this case the reactions are of birth, annihilation,
and death type. From  (\ref{tausc}) we find
\begin{equation}
\tau=\frac{2}{\gamma}\sqrt{\frac{\pi}{N}}
\frac{R_{0}^{3/2}}{(R_{0}-1)^{2}
(R_{0}+1)^{1/2}}
\exp\left\{2N\left[\left(1+\frac{1}{R_{0}}\right)
\ln\left(\frac{1+R_{0}}{2R_{0}}\right)+
1-\frac{1}{R_{0}}\right]\right\},
\label{ta1d2}
\end{equation}
recovering the result already obtained in \cite{AsMe10}.
In Fig.~\ref{fig:f1011} we plot the cases $a=d=1$
and $a=1$, $d=2$. The fact that for $d=2$ we
have annihilation, rather than competition as for $d=1$,
reduces the MET as one would expect. If we take $R_0$ to
infinity and compare the cases of $d=2$ and $d=1$,
both with $a=1$,
we find $\tau(d=2)/\tau(d=1)\sim \exp(-2N\ln2)$.
For both panels in Fig.~\ref{fig:f1011} we observe that
the MET increases very fast with the basic reproductive
number $R_0$. The comparison between
simulations and analytic results are in general
good, except when $R_0$ tends to the
critical value given in (\ref{cc}).
Indeed, the WKB theory breaks down if
the barrier $\Delta S$ tends to zero.
This happens if $p_f$ tends to 0. Considering (\ref{eqtr}),
the limit $p_f\rightarrow 0$ implies
$R_0\rightarrow 1/a$.
\end{widetext}

\section{Conclusions}
We have adopted an individual-based formulation to
describe the random dynamics of finite-sized populations.
Specifically, we have analyzed in detail
various possible microscopic
scenarios that all give
rise to the same macroscopic population-level
model, namely the Verhulst or
logistic population growth equation.
We have shown that
for birth and competition interactions between individuals,
$\text{X}\xrightarrow{\lambda} (a+1)\,\text{X}$,
$2\,\text{X}\xrightarrow{\mu} X$,
the population does not become
extinct, regardless of the value of the parameters.

If competition leads to annihilation of
the competitors, $\text{X}\xrightarrow{\lambda} (a+1)\,\text{X}$,
$2\,\text{X}\xrightarrow{\mu}
\emptyset$,
the ultimate fate of the population depends on
whether the
kinetics is parity conserving or not. The
parity of the total number of particles is preserved
in the even-offspring case.
This implies that
the population persists if $a$ is
even and $n_0$ is odd, because the
absorbing state is inaccessible.
On the other hand, if $a$
and $n_0$ are both even or if $a$ is odd,
the absorbing state is accessible and
the population becomes
extinct. It is worth noting that
these kinetic
rules
can be implemented as dynamical lattice models or
interacting particle systems, for example
as a contact process or a
branching-annihilating
random walk (BARW)
\cite{TaTr92,Je93a,Je93,ZhAv95,CaTae96,BaCa03,Li05}
and that parity conservation, or the lack thereof,
also plays a crucial role in the dynamics of
these spatially extended systems.
They can display
a nonequilibrium transition from a
nontrivial fluctuating steady state to an
absorbing state with no fluctuations
\cite{CaTae96}. This transition belongs to
different universality classes for parity-conserving and
nonparity-conserving models
\cite{Hi00}. The most prominent member of the first
class is the BARW with an even number of offsprings.
The dynamics of BARWs with even and odd number of offsprings
have been analyzed in detail in \cite{AvLeRe94}.

For those cases
where the population persists,
we have obtained analytic expressions for
the generating function and the PDF in the stationary state.
In particular, we have
determined the mean of the PDF and its coefficient
of variation.
For those cases where the population becomes extinct,
we have calculated the MET and have explored its dependence on
the microscopic
parameters. All our analytical results have been compared with
numerical simulations,
showing good agreement.

Our results provide
further evidence for the advantages of individual-based models.
They demonstrate that the microscopic
details of random events at the level of the individuals
lead to differences in the behavior of the system at
the population level. In the case that the population persists,
the characteristics of the stationary PDF
depend on the features of
the microscopic model. To illustrate this fact, we have focused
on the coefficient of variation.
Our results show, see Figs.~\ref{fig:f3} and \ref{fig:f5},
that an increase in $a$, the number of offsprings, and $d$,
the number of individuals removed due to competition, leads
to an increase in the variability of the population for the same
value of the macroscopic parameter, the carrying capacity $N$.
Measuring the coefficient of variation of a population
for a given value of $N$
provides therefore a means of
drawing inferences about the microscopic
details of the birth and competition processes.

Similarly, in the case that the population
becomes extinct, the MET depends sensitively on
the microscopic details of the model, as illustrated
by Figs.~\ref{fig:f67}, \ref{fig:f89}, and \ref{fig:f1011}.
For example, for the model
$\text{X}\xrightarrow{\lambda} (a+1)\,\text{X}$,
$2\,\text{X}\xrightarrow{\mu}
\emptyset$, we find that if $n_{0}$ and $a$ are even, then
the MET becomes significantly larger for the same carrying
capacity as $a$ increases, see Fig.~\ref{fig:f67}.
In contrast, the MET becomes significantly smaller for the
same carrying capacity as $a$ increases if $a$ is odd,
see Fig.~\ref{fig:f89}.
Extinction is always the ultimate fate
for the birth-competition-death model,
$\text{X}\xrightarrow{\lambda} (a+1)\,\text{X}$,
$2\,\text{X}\xrightarrow{\mu}
(2-d)\,\text{X}$, $\text{X}\xrightarrow{\gamma}
\emptyset$.
Figure~\ref{fig:f1011} demonstrates strikingly
the sensitive dependence of the MET on the
microscopic details of the system. Measuring
the MET for laboratory populations with a given basic
reproductive number provides therefore a means of
drawing inferences about the microscopic
details of the birth, death, and
competition processes.
Our results also imply that assessing the
extinction risks and survival times of natural
populations requires an understanding of the
microscopic details of the processes
that occur in
the system and should not be based
solely on phenomenological
models.

There are many other possibilities that
lead to the logistic
equation, for example if we
consider two different birth reactions
simultaneously, such as,
$\text{X}\xrightarrow{\lambda} 2\,\text{X}$ and
$\text{X}\xrightarrow{\lambda} 3\,\text{X}$.
Another possibility consists in
considering schemes with
four or even more reactions.
All these situations can be
analyzed in the same manner and with
the same techniques as used here.
A further intriguing possibility that deserves study are
reactions schemes where the number of offsprings, $a$, and
the number of individuals eliminated by exclusive competition,
$d$, fluctuate randomly between several values.

\begin{acknowledgments}
This research has been supported  (VM, DC)
the Ministerio de Ciencia e Innovaci{\'o}n under Grant
No. FIS2012-32334.
\end{acknowledgments}

\vfill

\bibliography{stoclog}

\end{document}